\documentclass[conference]{IEEEtran}
\usepackage{graphicx, epstopdf}
\usepackage{amssymb, amsmath}
\usepackage{commath}
\usepackage{setspace}
\usepackage{acronym}
\usepackage{mathrsfs}
\usepackage{ifthen}
\usepackage{textcomp}
\usepackage{float}
\usepackage{subfigure}
\usepackage{color}
\usepackage{cite}

\IEEEoverridecommandlockouts

\begin{document}
\bstctlcite{ICC09_Ref2:BSTcontrol}

\title{A Molecular Communication Link for Monitoring in Confined Environments}

\author{Song Qiu, Weisi Guo, Siyi Wang\textsuperscript{\dag}, Nariman Farsad\textsuperscript{\S}, Andrew Eckford\textsuperscript{\S} \\
School of Engineering, University of Warwick, United Kingdom \\
\textsuperscript{\dag}Institute for Telecommunications Research, University of South Australia, Australia \\
\textsuperscript{\S} Dept. of Electrical Engineering and Computer Science, York University, Canada \\
Corresponding Author Email: weisi.guo@warwick.ac.uk}

\maketitle

\begin{abstract}
In this paper, we consider a molecular diffusion based communications link that can reliably transport data over-the-air.  We show that the system can also reliably transport data across confined structural environments, especially in cases where conventional electromagnetic (EM) wave based systems may fail.  In particular, this paper compares the performance of our proprietary molecular communication test-bed with Zigbee wireless sensors in a metal pipe network that does not act as a radio wave-guide.  The paper first shows that a molecular-based communication link's performance is determined primarily by the delay time spread of the pulse response.  The paper go on to show that molecular-based systems can transmit more reliably in complex and confined structural environments than conventional EM-based systems.  The paper then utilizes empirical data to find relationships between the received radio signal strength, the molecular pulse spread, data rate (0.1 bits/s) and the structural propagation environment.
\end{abstract}

\section{Introduction}

Modern human civilization is partly defined by the intelligent and large-scale conversion of natural resources to useful fuels and tools.  The mining industry is the origin of this vital economic process and the global market capitalization of mining alone is close to \$1 trillion. A large part of mining and the processing of minerals occurs in subterranean or confined industrial locations.  Typically these locations are fraught with multiple natural and industrial hazards.  Therefore, there is a need to deploy wireless sensor networks to monitor events (i.e., structural integrity, dangerous chemicals, and health of workers) \cite{Wassell10, Forooshani13}.  Wireless communication and sensory equipment are constrained by electrical safety, hostile propagation environment, electromagnetic (EM) noise in these environments.  The challenge in subterranean (i.e., tunnels and mines) and industrial confined environments (i.e., pipe networks) is that the propagation pathloss for EM waves can be very high.  For example, in tunnels and underground mines, the pathloss distance exponent can be greater than 4 for a wide range of radio frequencies (0.9 to 2.4 GHz) \cite{Forooshani13}.  When the pipe or tunnel network can not act as a wave-guide for the data bearing EM wave, the pathloss is even greater \cite{Harvey55}.  There are a number of reasons why the structural environment can not act as a natural wave-guide to wireless carrier waves.  This could be due to the fact that many pipes and tunnels are not of constant dimension and are therefore cannot act as a good wave-guide, or that it is not always easy to generate the necessary carrier frequency for cost and physical dimension reasons.
\begin{figure}[t]
    \centering
    \includegraphics[width=1.00\linewidth]{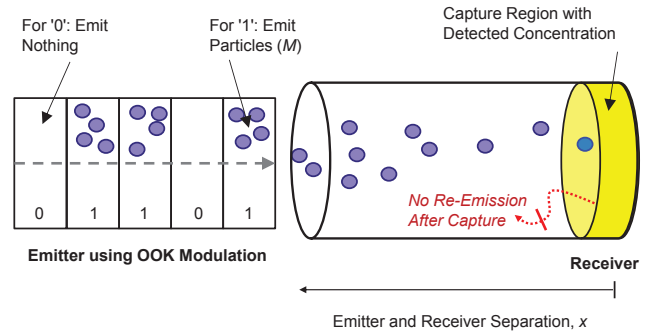}
    \caption{Illustration of molecular communication link using On-Off-Keying (OOK) modulation scheme with a receiver that has a capture zone.}
    \label{System}
\end{figure}
The motivation behind using molecules to carry general information lies trying to wirelessly communicate sensory information in the aforementioned environments.  In many industrial applications, wireless sensors are distributed in embedded locations. Such locations restrict the level of human access, antenna size and EM wave propagation. For such sensor applications, there is often a requirement to design small sized sensors that can deliver data without tether and at very low energy levels. Examples include monitoring corrosion in structures, and also in areas where one wants to minimise electromagnetic radiation or suffer from excess radiation interference.  This paper will show how sensors that demand a low data rate and have a high delay tolerance are especially suited to molecular communication channels.

Chemical based communications occurs on a body-to-body level (macro-scale) and on a cell-to-cell level (micro-scale).  In chemical communications, certain properties of chemical particles are used to modulate data, i.e., chemical mixture, concentration, or time of arrival.  In this paper, we utilize general chemical communications, whereby pulse modulated chemical pulses can represent any message, as opposed to pre-defined fixed singular messages.  Our earlier work has shown that we can encode any message reliably using an on-off-keying (OOK) system~\cite{Farsad13PLOS}.  In fact, this was the world's first over-the-air molecular communication system that can reliably transmit general messages.  The idea of communicating using chemicals has been around for a decade, and most work has been of a theoretical nature~\cite{Atakan07, Pierobon10, Eckford12}.  There has been some work investigating chemical signalling to control robots or mimic insect signalling, but the messages have been confined to specific commands encoded with specific chemical mixtures.  The paper is organised as follows.  In Section II, we recap on the importance of the pulse-response's time-spread in pulse modulated molecular systems and analyze the importance of time-spread.  In Section III, we outline our propagation experiments in complex structural environments.  In Section IV, we present our empirical results and derive propagation models for future reference.

\section{Pulse Modulated Molecular Communications}

Pulse modulated molecular communications is not known in nature, where organisms predominantly use chemical mixtures to encode data.  Whilst it is possible to encode over multiple chemical compounds (prevalent in nature), we focus on a single compound and note that multiple compounds will scale the capacity linearly in a bandwidth concept.  Pulse-modulation is relatively easy to implement and an illustration of our system is presented in Fig.~\ref{System}, whereby an emitter utilizes on-off-keying (OOK) to encode and transmit data across a pipe network (semi-infinite space).  The receiver captures the data, ensuring that molecules are converted to electronic charge and no re-emission of molecules from the receiver is possible.  In such a system, errors at the receiver can arise from: i) noise and chemical interference from the environment and hardware, ii) inter-symbol-interference (ISI), and iii) changes to the propagation environment (i.e., air flow, temperature, humidity).  The most influential source of error arises from ISI and changes to the propagation environment.  This is because unlike EM radio waves, molecules diffuse in a more stochastic manner, and the time of arrival of a pulse (delay spread) has a very large variance.
\begin{figure}[t]
    \centering
    \includegraphics[width=1.00\linewidth]{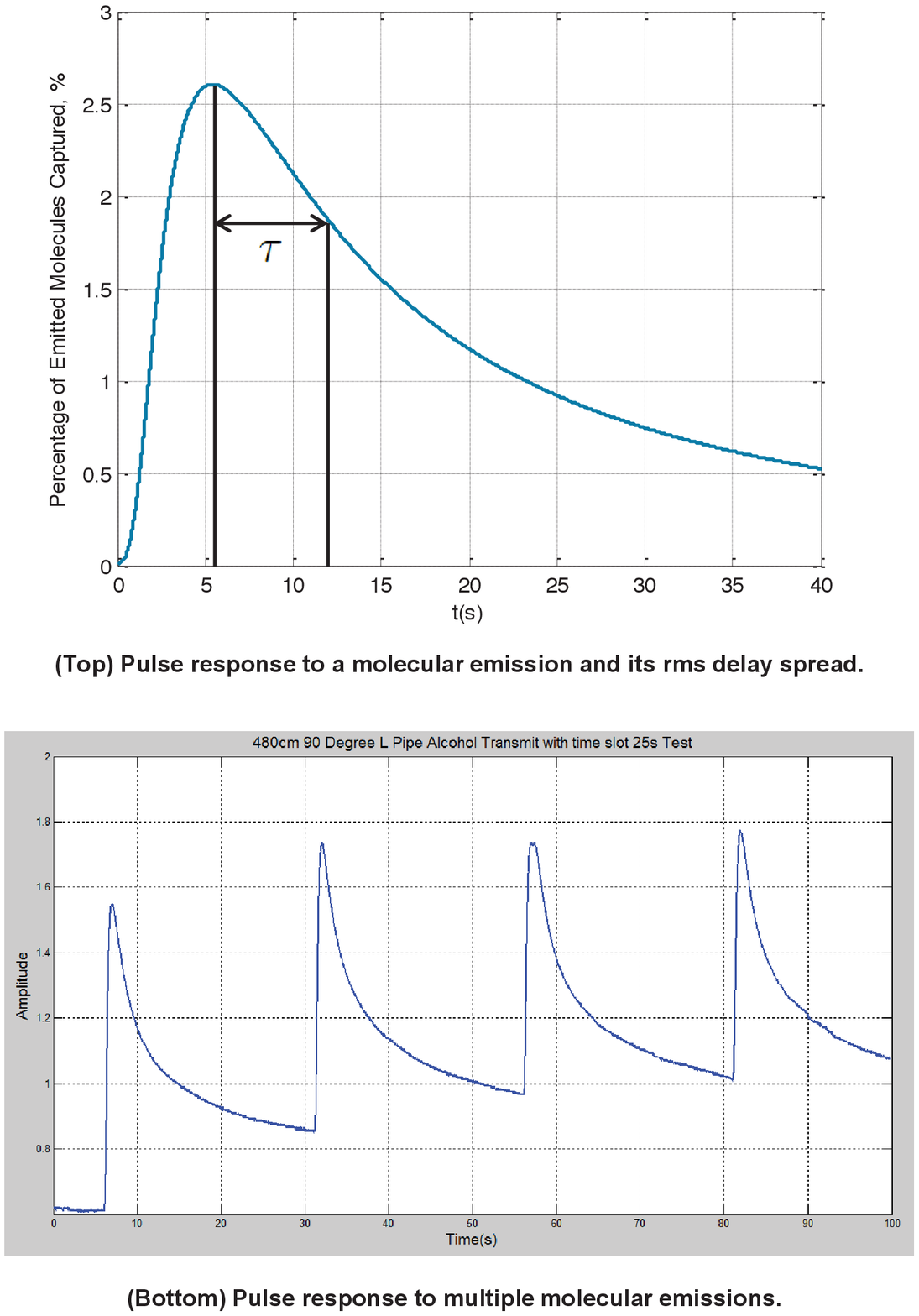}
    \caption{(Top) Received impulse response and corresponding delay spread of a molecular transmission pulse.  Data for $x=10$m of distance with ambient drift currents. (Bottom) Received impulse response of multiple molecular transmission pulses.  Data for $x=4.8$m of L-shaped pipe network.}
    \label{DelaySpread}
\end{figure}
\begin{figure*}[t]
    \centering
    \includegraphics[width=0.65\linewidth]{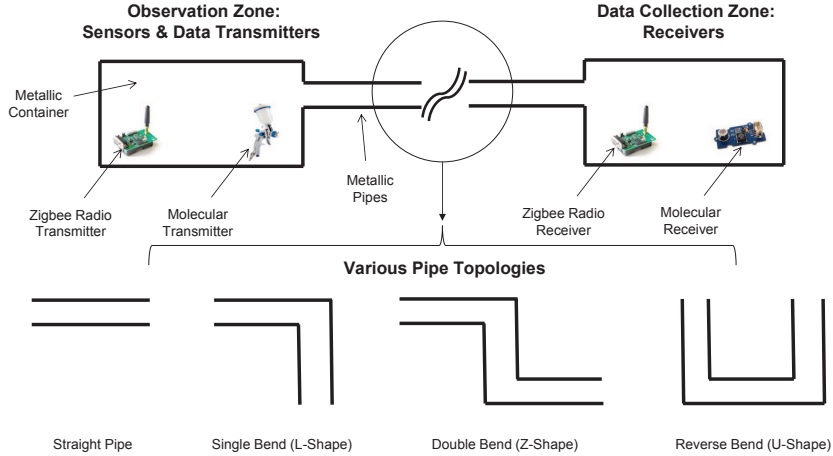}
    \caption{Illustration of propagation environment consisting of 2 metallic boxes connected by a metallic pipe with various lengths and bends.}
    \label{Experiment}
\end{figure*}

To demonstrate the concept, we consider the transmitter emitting a series of impulses, each containing $M$ molecules.  Each impulse input (spray) is defined as a rectangular pulse with a finite small duration (approx. 500ms).  The molecular concentration (impulse response) at a distance $x$ and at the time $t$ after emission is found using hitting probability equations that satisfy the Fokker-Planck equation \cite{Ross96}:
\begin{equation}
\label{eqn:hit}
\phi_{\text{hit}}(x,t) = \frac{\mathsf{M}}{\sqrt{\pi D t}} \exp{\left[-\frac{(x-vt)^{2}}{4 D t}\right]},
\end{equation} for a positive laminar air current of velocity $v$, and a diffusivity $D$.  This is a familiar expression used in most theoretical works in molecular communications~\cite{Atakan10}.  When the receiver captures molecules, the full derivation with drift velocity is complex.  However, when zero-drift velocity is consider, the number of captured molecules over a receiver sample time of $\tau$ at time $T$ is a difference equation \cite{Guo13EL}:
\begin{equation} \begin{split}
\label{eqn:capture}
\phi_{\text{capture}}(x,T,\tau) = \mathsf{M} \mathrm{erfc}\left(\frac{x}{2\sqrt{Dt}}\right)|_{T}^{T+\tau}.
\end{split} \end{equation}

In Fig.~\ref{DelaySpread}(top), for an impulse input (burst of molecules emitted), the delay spread at the receiver is shown.  There is a linear conversion between the number of molecules captured and the resulting electrical current induced in the circuits.  In this paper, the authors define the delay spread as the peak-to-3dB power point of the response.  That is to say from the peak of the voltage amplitude to the $\sqrt{2}$ of the amplitude:
\begin{equation} \begin{split}
\label{eqn:spread}
\tau = t_{\text{max}[\phi]} - t_{\text{rms}[\phi]}.
\end{split} \end{equation}
Given that the large delay spread will cause ISI to a large number of future symbols, ISI will dominate the reliability and rate of the communications link.  In Fig.~\ref{DelaySpread}(bottom), a sequence of molecules is emitted showing how the delayed response of previous pulses causes an inflation of the response in subsequent pulses.  Therefore, characterizing the magnitude of the delay spread in molecular communications, as a function of the propagation environment's parameters is more essential than the signal-to-noise ratio (SNR).  From Fig.~\ref{DelaySpread}(bottom), there is actually very little in terms of additive noise on the receiver side, and the predominant source of error is ISI.
\begin{table}[t]
    \caption{Test Equipment Parameters and Data}
    \begin{center}
        \begin{tabular}{|l|l|}
          \hline
          \emph{Feature}                    & \emph{Parameter and Data}         \\
          \hline
          \textbf{EM Radio System}          & \textbf{Zigbee}                   \\
          Zigbee Module                     & Telegesis ETRX357                 \\
          Transmission Band                 & 2.4 GHz                           \\
          No. of Channels                   & 16                                \\
          Transmit Power                    & 1 mW                              \\
          Receiver Sensitivity              & -99 dBm                           \\
          Peak Antenna Gain                 & 2.1 dBi                           \\
          \hline
          \textbf{Molecular System}         & \textbf{Kinboshi}                 \\
          Alcohol Sensor                    & MQ3                               \\
          Chemical Carrier                  & Alcohol                           \\
          Chemical Sensitivity              & Alcohol vs. Air (30--600$\times$) \\
          Temp. Sensitivity                 & Peak to RMS (-10 to 12 C)         \\
          Humidity Sensitivity              & 13\% (RH: 85\% to 33\%)           \\
          Response                          & Linear                            \\
          Air Flow                          & Positive Turbulent (14 m/s)       \\
          \hline
          \textbf{Propagation Environment}  & \textbf{2 Tanks \& Pipe Network}  \\
          Pipe Thickness                    & 2 mm                              \\
          Pipe Diameter (Hollow)            & 4 cm                              \\
          Waveguide Band                    & approx. 4-6 GHz \cite{Harvey55}   \\
          Pipe Bend                         & 90 degrees                        \\
          Iron Container Volume             & 3600 cm$^{3}$                     \\
          Box and Pipe Material             & Iron                              \\
          \hline
        \end{tabular}
    \end{center}
    \label{Parameters}
\end{table}

\section{Experimental Setup}

\subsection{Propagation Environment}

This paper primarily focuses on the characterization of the relationship between ISI and the delay spread, and how the propagation environment affects this.  Our experiments assume a constant humidity and temperature, and only vary the structural topology of the environment.  Figure~\ref{Experiment} illustrates the experimental setup between an observation zone and a data collection zone.  Each zone consists of a metal tank, interconnected by an iron pipe network.  The two tanks are defined as two zones:
\begin{enumerate}
  \item In the \emph{observation zone}, we assume there is an event of interest (e.g., state of contents in a storage tank, structural integrity of the tank itself), and there are sensors to detect the event and report the data wirelessly.
  \item In the \emph{data collection zone}, there are receivers that await the data reported from the observation zone.
\end{enumerate}
The pipe network is a flexible design, whereby the length of individual pipe sections and the number of bends can be adjusted.

The challenge in many industrial applications is that the metallic tanks are connected by complex pipes (e.g., ventilation pipes) and EM waves do not necessarily propagate well through complex pipe technologies.  This is especially the case, when the pipes can not act as a wave-guide to the data bearing EM-waves.  This could be because there are constraints to what EM frequency bands are available for use and restrictions on the antenna dimensions required to generate the appropriate waves.  As shown in Table~\ref{Parameters}, we consider EM-waves (2.4 GHz) travelling outside (below the lower cut-off frequency) the wave-guide bandwidth of the pipe (4-6 GHz) \cite{Harvey55}.  Future research will examine in-band performance comparisons.  Alternative solutions include drilling holes to create an improved EM propagation path between the containers, or deploying a wired communication system.  Drilling holes through the tanks is not an attractive solution as the tanks can be filled with fluid or gas contents, or the holes can compromise the tank's function.  A wired solution is not attractive as it requires prior infrastructure deployment in the pipe network and can cause blockages forming in the long run.  Therefore, data needs to be communicated through the pipes without wires.  The paper aims to find an empirical relationship between the length of the pipes, the number of bends, and the communication capability of EM-based and molecular-based systems.

\subsection{Communication System Hardware}

In the EM-based communication system, we use a standard Telegesis ETRX357 Zigbee module. The transmit power is approximately 1mW and the receiver sensitivity is -99dBm, which can deliver a bit-error-rate below 1\%.  The transmission band is 2.4 GHz and the antenna shape is approximately homogeneous, with a peak gain of +2.1 dBi and a mean gain of -1.2 dBi.  In the molecular-based communication system, we use the design known as Kinboshi, which is detailed in our previous work \cite{Farsad13PLOS}.  The sensitivity level is available for different chemicals, temperatures (Celsius), and relative humidity (RH) levels.  It is approximately linear for all concentration values.  The data for this particular sensor shows that ideal molecular communications performance requires the transmitter to use alcohol molecules and an environment that has a low humidity and low temperature (-10 to 12C).  Furthermore, a positive (turbulent) air current drift is induced to accelerate the rate of diffusion.  The specific encoding and decoding of information in the Kinboshi molecular system is beyond the scope of this paper and can be found in the seminal work \cite{Farsad13PLOS}.

\begin{table}[t]
    \caption{Summarized Results Data: mean value from 3--4 independent test batches with standard deviation in brackets}
    \begin{center}
        \begin{tabular}{|l|l|l|}
          \hline
          Structure Shape               & Radio Signal          & Delay Spread        \\
          \hline
          \textbf{Baseline}             &                       &                     \\
          Free Space (4.0m)             & -70 dBm (1)           & 17s (3)             \\
          1 Sealed Tank (2.5m)          & No Signal             & No Signal           \\
          2 Sealed Tanks (1.0m)         & No Signal             & No Signal           \\
          2 Open Tanks  (4.0m)          & -90 dBm (1)           & 65s (11)            \\
          \hline
          \textbf{Pipeline Shape}       &                       &                     \\
          Straight (0 bend, 3.6m)       & -79 dBm (1)           & 3.57s (0.4)         \\
          L-Shape (1 bend, 3.7m)        & -92 dBm (0)           & 4.29s (0.3)         \\
          L-Shape (1 bend, 4.8m)        & No Signal             & 6.24s (0.5)         \\
          Z-Shape (2 bends, 3.8m)       & -93 dBm (3)           & 9.07s (2.0)         \\
          U-Shape (2 bends, 3.9m)       & No Signal             & 8.81s (1.1)         \\
          \hline
          \textbf{Pipeline Length}      &                       &                     \\
          Short (0 bend, 1.3m)          & -62 dBm (1)           & 2.20s (0.2)         \\
          Medium (0 bend, 2.5m)         & -71 dBm (1)           & 2.91s (0.4)         \\
          Medium (1 bend, 2.6m)         & -87 dBm (1)           & 4.45s (0.4)         \\
          Long (0 bend, 3.6m)           & -80 dBm (1)           & 3.57s (0.4)         \\
          Long (2 bends, 3.9m)          & No Signal             & 8.81s (1.1)         \\
          L-Shape (1 bend, 4.8m)        & No Signal             & 6.24s (0.5)         \\
          \hline
        \end{tabular}
    \end{center}
    \label{Data}
\end{table}

\begin{figure}[t]
    \centering
    \includegraphics[width=1.00\linewidth]{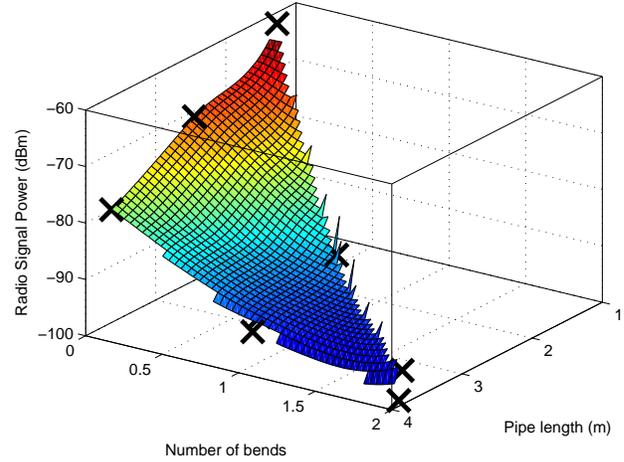}
    \caption{Radio signal power loss (dB) vs. different pipe lengths (m) and number of pipe bends.}
    \label{ResultsRadio}
\end{figure}
\begin{figure}[t]
    \centering
    \includegraphics[width=1.00\linewidth]{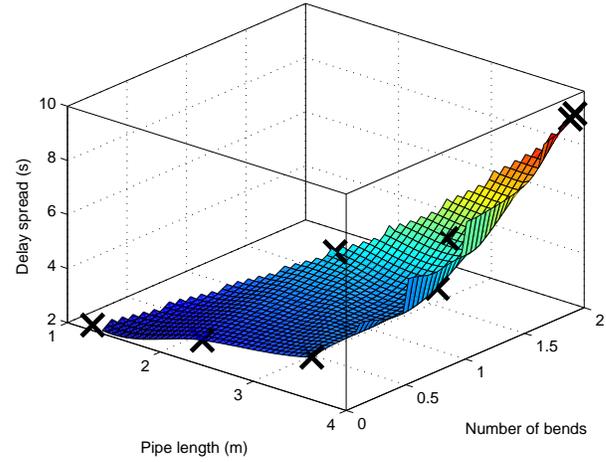}
    \caption{Molecular signal time spread (s) vs. different pipe lengths (m) and number of pipe bends.}
    \label{ResultsNano}
\end{figure}
\begin{figure*}[t]
    \centering
    \includegraphics[width=0.95\linewidth]{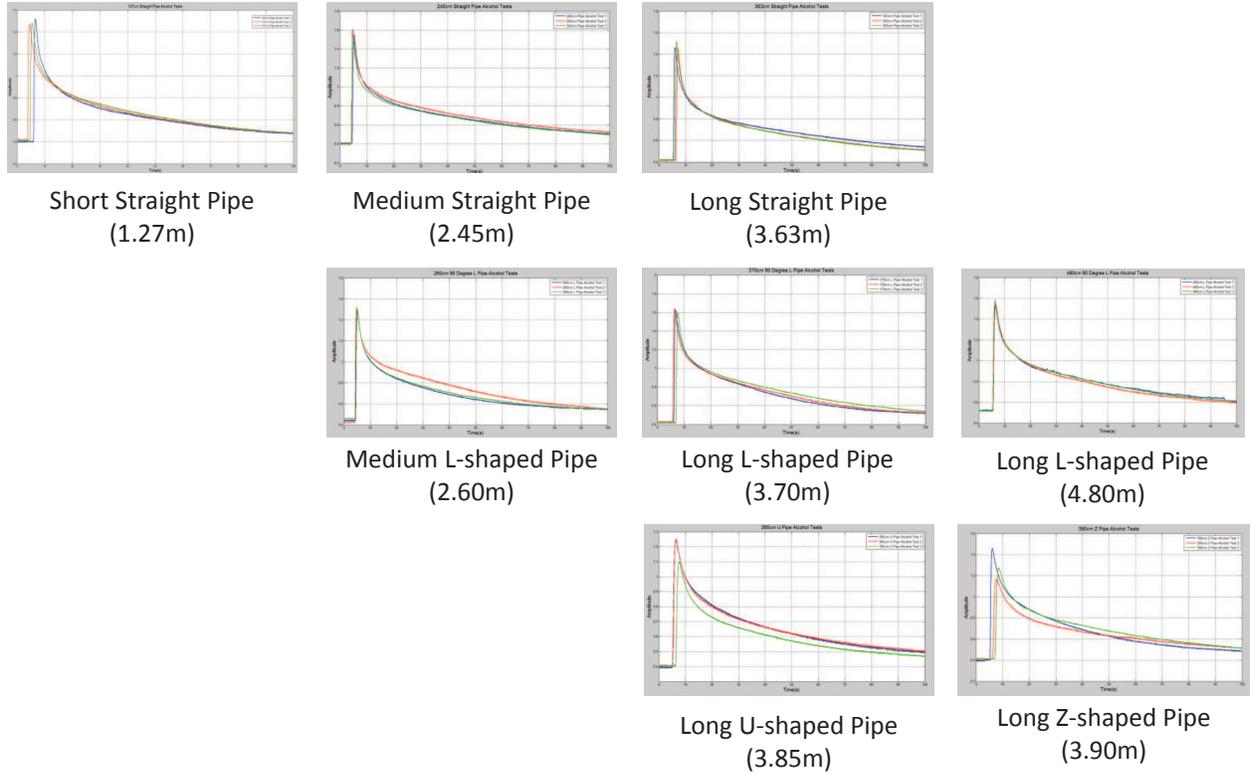}
    \caption{Full set of molecular impulse responses for different pipe configurations and dimensions.  Each set of result contains 3 independent test data trials.}
    \label{Full_Results}
\end{figure*}

\section{Results}

\subsection{Propagation Analysis}

The paper first presents the summarized radio and molecular communication results in Table~\ref{Data}, which shows the significant propagation results for consideration.  The baseline results show that (as expected) neither the radio nor the molecular signal can penetrate sealed metallic tanks.  However, communications is possible in both the free-space and open tanks cases.  For molecular communications, the delay spread in an open or semi-open environments is very high.  Intuitively, this is due to the fact that molecules carrying signals can diffuse anywhere and the arrival process at the receiver is highly stochastic.  These numbers can be used a reference point for comparison with more complex structural topologies.  The full set of impulse response shapes for different pipe dimensions and configurations can be found in Fig.~\ref{Full_Results}.  It can be observed that the shape is very much similar, independent of the pipe network, and only the delay spread varies.  The variance is also very small across all experimental data.

When the tanks are sealed, but joined by a metallic pipe structure (Fig.~\ref{Experiment}), the results are of more interest.  The first observation is that the pipe shape (e.g., U-, or Z-Shape) doesn't strong affect the performance of either the radio or the molecular communication link.  That is to say the orientation of the bend is not important.  However, the number of bends does strongly affect the molecular communication link.  In the second part of Table~\ref{Data}, data is presented for 0 to 2 bends for a similar overall distance.  The delay spread of the molecular pulse response dramatically increases with each increased bend.  In the third part of Table~\ref{Data}, data is presented for different pipe lengths.  It shows that the radio performance is strongly affected by both the pipe length and the number of bends (especially the first bend), whereas the delay spread of molecular communications is only strongly correlated with the number of bends.  It can be seen that the results standard deviation is generally small for the radio experiments (mostly 1 dBm or less), and remain small for most molecular experiments (mostly 1s or less).  With greater number of pipe bends and distances, the radio and molecular results' standard deviation can rise rapidly to 3 dBm and 2s respectively.

An approximate propagation loss relationship for radio communications in this proprietary experiment (consisting of 2 metal tanks connected with iron pipes) is:
\begin{equation} \begin{split}
\label{eqn:Loss}
\text{PL}_{\text{radio}} =
\left\{
\begin{array}{l l}
-8 \text{ dB/m} & \mbox{straight pipe} \\
-10 \text{ dB/bend} & \mbox{first bend} \\
\end{array} \right..
\end{split} \end{equation}

An approximate delay spread relationship for molecular communications in this proprietary experiment (consisting of 2 metal tanks connected with iron pipes) is:
\begin{equation} \begin{split}
\label{eqn:Loss}
\tau =
\left\{
\begin{array}{l l}
+0.6 \text{ s/m} & \mbox{straight pipe} \\
+35 \text{ to } 50\% \text{ increase/bend} & \mbox{pipe network} \\
\end{array} \right..
\end{split} \end{equation}

This conclusion can be visualized by the 3D plots of radio and molecular performance as a function of the number of pipe bends and the overall pipe length.  In Fig.~\ref{ResultsRadio}, the plot shows how the signal power loss (dB) varies with different pipe lengths and number of bends strongly.  For the number of bends, it is significant for only the first bend.  On the other hand, the results in Fig.~\ref{ResultsNano} show that the delay spread dramatically increases with each bend in a non-linear fashion.  The key conclusion are as follows:
\begin{itemize}
  \item Molecular communications can be established in scenarios where a reliable radio-link can not be established due to propagation losses in complex confined environments.  In our case, it was for long pipe lengths (over 4m) and a high number of bends (2 or over);
  \item Radio communications is more sensitive to the pipe length and the first pipe bend, whereas molecular communications is the more sensitive to the number of bends.
\end{itemize}
Generally speaking, molecular communications is more resilient than radio communications in the short ranged complex topologies investigated in this paper.  Admittedly, radio communications can be improved with frequency-hopping, greater transmit power, and even relaying technologies.  However, the paper has used a simple experiment to demonstrate the potential of a simple proprietary molecular system with no complex modulation or coding mechanism, over an established commercial radio system.
\begin{figure}[t]
    \centering
    \includegraphics[width=1.00\linewidth]{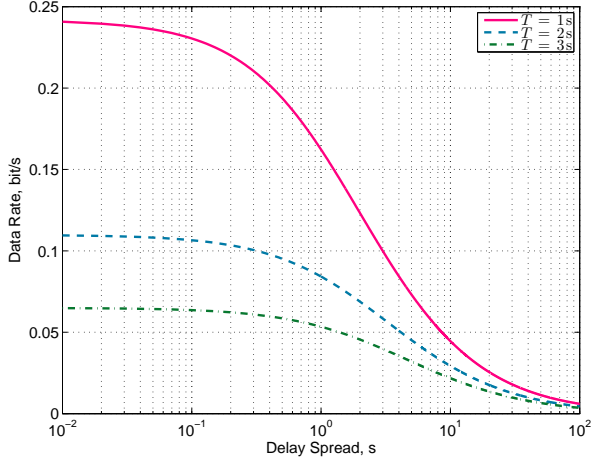}
    \caption{Achievable data rate of molecular communication link as a function of the delay spread $\tau$, and the arrival delay $T$.}
    \label{Capacity}
\end{figure}

\subsection{Bit Error Rate and Throughput Analysis}

Recall that the delay spread is defined in \eqref{eqn:spread} as the duration under which the received power falls from the peak to the r.m.s. point.  Let us assume that some factor of the delay spread is also a reasonable sampling period at the receiver to capture most of the energy induced by the transmitted molecules.  Therefore, it is sufficient to say any molecules outside this region will contribute to ISI to symbols before and after it.  Thus, one can deduce the bit error rate (BER) to be \cite{Guo13EL}:
\begin{equation} \begin{split}
\label{eqn:error}
p_{\text{molecular}}(x,T,n,\tau) = 1 - \mathrm{erfc}\left(\frac{x}{2\sqrt{Dt}}\right)|_{T}^{T+n\tau},
\end{split} \end{equation} where $n$ denotes some positive multiplier to the delay spread, and $T$ is the arrival time of the peak response.  In order to ensure a BER below a certain value, the sampling duration $n\tau$ needs to be increased.  A high delay spread will cause a greater sampling duration and hence a slower rate of transmission (lower throughput).  In molecular communications with a non-adaptive OOK modulation, the throughput (good-put) is the number of successful bits delivered \cite{Guo13EL}:
\begin{equation} \begin{split}
\label{eqn:rate}
R_{\text{molecular}}(x,T,n,\tau) = \frac{\mathrm{erfc}\left(\frac{x}{2\sqrt{Dt}}\right) |_{T}^{T+n\tau}}{n\tau}  \quad \text{bits/s/chemical}.
\end{split} \end{equation}   As shown in Fig.~\ref{Capacity} the throughput is in the order of 0.2--0.3 bits/s for a typical molecular communication link spanning several metres at different delay spread time and sampling delay $T$ values.  These capacity values are in line with experimental results derived by us and by our collaborators independently \cite{Farsad13PLOS, Farsad13JSAC}.  No doubt, more advanced modulation schemes involving chemical mixtures can achieve higher rates.  However, given that the application is sending low rate sensory data, the key issues in research should not be focused purely on channel capacity.

\section{Conclusions}

This paper has used a configurable structure to demonstrate that electromagnetic (EM) based communication links can be unreliable, when a pipe network can not act as a wave-guide. However, recently prototyped molecular based communication systems can slowly, but reliably transport data.  The paper argues that whilst most systems are concerned with signal-to-noise ratio levels, molecular-based systems are primarily concerned with the delay spread of the impulse response ($\tau$).  Using empirical measurement results, the paper shows that the delay spread is more strongly affected by the number of structural bends than the distance of transmission.  The paper shows that given a certain bit-error-rate target, the larger the delay spread, the greater the receiver sampling duration needs to be, which lowers the data rate.  In general the data rate is in the order of 0.1 bits/s for a simple OOK modulated system.   \\

~~~~~~~~~~~~~~~~~~~~~~~Acknowledgement\\
The work in this paper is partly supported by the Royal Society International Exchange Scheme, University of Warwick's Global Research Priorities (GRP), and the Australian Research Council (ARC).

\bibliographystyle{IEEEtran}
\bibliography{IEEEabrv,Ref}
%%%%%%%%%%%%%%%%%%%%%%%%%%%%%%%%%%%%%%%%%%%%%%%%%%%%%%%%%%%%%%%%%%%%%%%%%%%%%%%

\end{document}